\documentclass[prb,twocolumn,superscriptaddress,bibnotes]{revtex4}
\usepackage{graphicx}

\begin{document}

\title{Ultrafast Insulator-Metal Phase Transition in VO$_2$ Studied by Multiterahertz Spectroscopy}

\author{A. Pashkin}
\author{C. K\"ubler}
\author{H. Ehrke} \altaffiliation{Present address: Max Planck Research Group for Structural Dynamics, University of Hamburg, CFEL, 22607 Hamburg, Germany}
\affiliation{Department of Physics and Center for Applied Photonics, University of Konstanz, 78457 Konstanz, Germany}

\author{R. Lopez}
\affiliation{Department of Physics and Astronomy and Institute of Advanced Materials, Nanoscience and Technology, University of North Carolina,
Chapel Hill, North Carolina 27599, USA} \affiliation{Department of Physics and Astronomy, Vanderbilt University, Nashville, Tennessee 37235,
USA}

\author{A. Halabica}
\author{R. F. Haglund, Jr.}
\affiliation{Department of Physics and Astronomy, Vanderbilt University, Nashville, Tennessee 37235, USA}

\author{R. Huber}
\author{A. Leitenstorfer}
\affiliation{Department of Physics and Center for Applied Photonics, University of Konstanz, 78457 Konstanz, Germany}

\begin{abstract}
The ultrafast photoinduced insulator-metal transition in VO$_2$ is studied at different temperatures and excitation fluences using multi-THz
probe pulses. The spectrally resolved mid-infrared response allows us to trace separately the dynamics of lattice and electronic degrees of
freedom with a time resolution of 40~fs. The critical fluence of the optical pump pulse which drives the system into a long-lived metallic state
is found to increase with decreasing temperature. Under all measurement conditions we observe a modulation of the eigenfrequencies of the
optical phonon modes induced by their anharmonic coupling to the coherent wave packet motion of V-V dimers at 6.1~THz. Furthermore, we find a
weak quadratic coupling of the electronic response to the coherent dimer oscillation resulting in a modulation of the electronic conductivity at
twice the frequency of the wave packet motion. The findings are discussed in the framework of a qualitative model based on an approximation of
local photoexcitation of the vanadium dimers from the insulating state.
\end{abstract}

\maketitle

\section{Introduction}

The insulator-metal transition in transition metal oxides, first reported by Morin,\cite{Morin59} is characterized by a sharp decrease in
resistivity by several orders of magnitude when the compound is heated above a critical temperature $T_\mathrm{c}$. Later a large number of
compounds have been found to exhibit this type of phase transition. Extensive experimental information has been collected and the physical
mechanisms responsible for the transition have been relatively well understood in many cases.\cite{Imada98} Nevertheless, the origin of the
phase transformation is still under debate in some systems. VO$_2$ is a classical example that has attracted considerable interest for
fundamental reasons as well as for possible applications, since its critical temperature $T_\mathrm{c}$ = 340~K (67~$^{\circ}$C) is located in
the vicinity of ambient conditions.

\begin{figure}
\centerline{\includegraphics[angle=0,width=0.85\columnwidth]{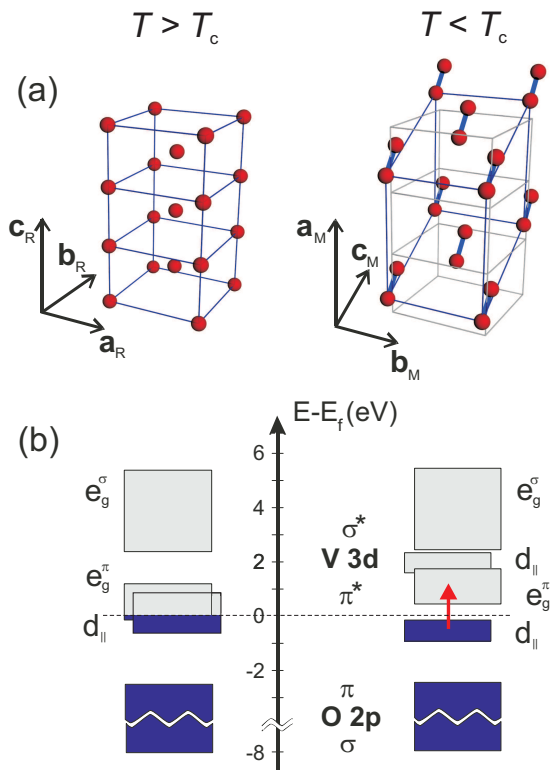}} \caption{(color online). (a) Vanadium sublattice of the rutile (left)
and monoclinic (right) crystal structure of VO$_2$. (b) Schematic band structure according to Refs.~\onlinecite{Goodenough71} and
\onlinecite{Koethe06}. The red arrow denotes the photoexcitation by an optical pump pulse with an energy of 1.5~eV.} \label{fig:structure}
\end{figure}

The change of the electronic properties in VO$_2$ is accompanied by a remarkable modification of the structure from the highly symmetric rutile
to the low-temperature monoclinic phase where tilted V-V pairs are formed.\cite{Morin59} The phase transition is of first order as confirmed by
the divergence of the heat capacity and the presence of latent heat.\cite{Berglund69} The high-temperature rutile phase of VO$_2$ (space group
D$_{4h}^{14}$) can be visualized in terms of a body-centered tetragonal lattice formed by the vanadium atoms as depicted in
Fig.~\ref{fig:structure}(a), with each metal atom surrounded by an oxygen octahedron.\cite{McWhan74} Each octahedron has virtually cubic
symmetry with very small orthorhombic distortions. Symmetry lowering during the transition into the monoclinic phase (space group C$_{2h}^{5}$)
leads to the doubling of the unit cell and the formation of V-V pairs\cite{Heckingbottom62,Longo70} as illustrated in
Fig.~\ref{fig:structure}(a). The soft mode of the rutile lattice corresponding to this structural transformation is located at the point $R$ =
(1/2,0,1/2) of the Brillouin zone.\cite{Brews70,Terauchi78,Gervais85} Correspondingly, the monoclinic phase can be mapped onto the rutile
lattice by two normal vibrational modes of vanadium atoms located at the $\Gamma$-point as a result of the Brillouin zone folding in the
monoclinic phase.\cite{Paquet80}

Working out the exact details of the electronic band structure of VO$_2$ poses a true challenge to modern many-body theories. In particular, the
correct prediction of the insulating gap remains a difficult task because the hierarchy of the microscopic degrees of freedom (spin, charge,
orbital and lattice) contributing to the phase transition is not clear. However, the phenomenological band scheme proposed by
Goodenough\cite{Goodenough60,Goodenough71} provides a quite realistic description by using simple arguments based on molecular orbitals. It
serves as a starting point for the discussion of the electronic properties of VO$_2$.

In metallic VO$_2$, hybridization between the oxygen $2p$ and vanadium $3d$ orbitals leads to $\sigma$- and $\pi$-type overlap.\cite{Eyert02}
The filled $\sigma$ and $\pi$ states will be primarily of O $2p$ character, whereas the corresponding antibonding bands are dominated by the V
$3d$ orbitals [Fig.~\ref{fig:structure}(b)]. The overlap of vanadium orbitals along the tetragonal $c_R$ axis parallel to the vanadium chains
results in the formation of a narrow band commonly labeled $d_{||}$. The Fermi level is located within all of the three $t_{2g}$ bands which are
still partially overlapping despite being no longer degenerate as shown in Fig.~\ref{fig:structure}(b).

Upon entering the low-temperature insulating phase, the experimentally observed structural changes bring the vanadium atoms closer to the apex
of the barely modified oxygen cage, thus increasing the hybridization. Now only the $d_{||}$ band is occupied and the dimensionality of the
electronic system is effectively reduced to unity. Moreover, photoemission experiments report a splitting of the $d_{||}$
band\cite{Shin90,Koethe06} by 2.5~eV that eventuates in the formation of a band gap of 0.6 - 0.7 eV between the lower $d_{||}$ band and the
$e_g$ band and the transition to the insulating state.\cite{Laad69}

Several theoretical models have been proposed based on Peierls\cite{Goodenough71,Wentzcovitch94} and Mott-Hubbard\cite{Zylberstejn75,Rice94}
scenarios. The former supposes a lattice instability caused by electron-phonon interaction within a system of reduced dimensionality. The latter
is based on strong electron-electron correlation effects. Recent \emph{ab initio} calculations on the basis of cluster dynamical mean-field
theory\cite{Biermann05} have demonstrated that the interplay of strong electronic interactions and the dimerization of vanadium atoms in the
low-temperature phase play a crucial role in forming the energy gap. In this scenario which may be described as a "many-body Peierls insulator",
strong Coulomb correlations in VO$_2$ lead to the formation of dynamical V-V singlet pairs that are necessary to trigger a Peierls transition.
The calculated electronic structure and optical conductivity in both phases of VO$_2$ agree well with experimental
results.\cite{Okazaki0406,Koethe06,Tomczak09}

The delicate balance of interactions drives transition metal oxides, such as VO$_2$, into a critical regime that is ruled by phase competition
and that reacts exceedingly sensitively to external stimuli. When ultrafast photoexcitation favors one of the competing phases via the
interaction of a photoexcited state with lattice, spin, or charge degrees of freedom, a dramatic phase conversion may occur. These phenomena are
highly cooperative so that the structural relaxation processes of the electronic excited states are not independent, as in conventional dilute
excitonic or photo-chemical absorption, but entail a photoinduced phase transformation toward a new lattice structure and electronic
order.\cite{Yonemitsu06,Yonemitsu08} This scenario opens the way for a light pulse to induce symmetry breaking from a stable phase and so to
establish a new long-range order. In this context, ultrafast time-resolved techniques are emerging as a tool to study phase transitions in
complex materials.\cite{Tokura06} Time is introduced as an additional parameter. It promotes the possibility to unravel the contributing degrees
of freedom and advance our understanding of sophisticated physical processes. Evidently, the time-resolved technique of choice should be
sensitive to the relevant degrees of freedom.

The interest in the insulator-metal transition of VO$_2$ was rejuvenated after Becker et al.\cite{Becker94} discovered that the metallic state
of VO$_2$ may not only be induced thermally but also optically on an ultrafast timescale. Later, X-ray diffraction\cite{Cavalleri01} and
optical\cite{Cavalleri04} pump-probe experiments proved the non-thermal character of the transition and suggested the limiting time for
switching to be less than 100~fs. A large transient pump-induced increase of the conductivity in the THz frequency range was recently reported
by several groups.\cite{Kubler07,Hilton07,Nakajima08} Therefore, nowadays VO$_2$ has attracted considerable attention due to potential
applications such as ultrafast control of light in VO$_2$-based photonic crystals,\cite{Mazurenko05} optical
switches,\cite{Soltani06,Cilento10,Kyoung10} and optical storage devices.\cite{Eden81} Efficient harvesting of this large technological
potential demands a thorough understanding of the microscopic physical processes.

Recently, we reported measurements of the insulator-metal transition in VO$_2$ induced by 12-fs optical pulses and probed by ultrabroadband
multi-THz transients.\cite{Kubler07} These experiments made it possible to discriminate spectrally between the excitation dynamics of electronic
and lattice degrees of freedom, revealing their fundamentally different character. Based on these observations a novel qualitative model of the
photoinduced insulator-metal transition in VO$_2$ has been suggested.

The present paper extends the previous short report by addressing such important aspects as a temperature dependence of the photoinduced changes
and a detailed insight into the coherent oscillation dynamics. In this paper, we present a systematic study of the photoinduced transition at
different temperatures and excitation fluences. The minimum energy of the pump pulse required to induce the cooperative non-thermal transition
into the metallic state is found to be comparable to the thermodynamic energy difference between the monoclinic and rutile phases. This fact
indicates an ultrafast switching of the lattice structure that instantaneously follows the electronic excitation of vanadium dimers. As a
characteristic fingerprint of this process a modulation of the eigenfrequencies of the optical phonon modes induced by their anharmonic coupling
to the coherent wave packet motion of V-V dimers is observed under all measurement conditions. In addition, a modulation of the electronic
response at twice the frequency of the wave packet motion gives evidence of a quadratic coupling to the structural order parameter. The observed
phenomena allow a deepened insight into the interplay of electronic and lattice degrees of freedom during the insulator-metal phase transition
in VO$_2$. The qualitative model of the ultrafast phase transition proposed previously is discussed in view of the additional results.

\section{Experiment}\label{experiment}

The sample is a 120~nm thin film of polycrystalline VO$_2$ grown by pulsed laser deposition on a CVD diamond substrate.\cite{Suh04} A
vanadium-metal target is ablated in 250 mTorr of oxygen by a KrF excimer laser (repetition rate 25 Hz) at a wavelength of 248 nm and a fluence
of 4 mJ/cm$^2$. The as-grown films are non-stoichiometric with an approximate composition close to VO$_{1.7}$. Therefore, the film is
subsequently oxidized at 450~$^{\circ}$C under $7\times10^{-3}$~mbar of oxygen pressure. After oxidizing to VO$_2$, the stoichiometry is
verified via Rutherford backscattering and observation of electrical and optical switching, while crystallinity and phase homogeneity are
confirmed by x-ray diffraction.

\begin{figure}
\centerline{\includegraphics[angle=0,width=1\columnwidth]{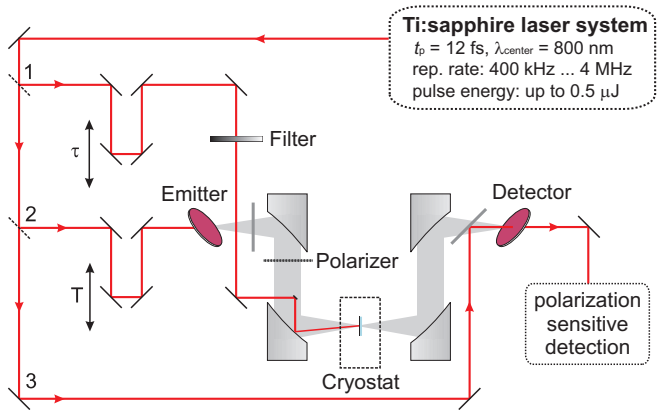}} \caption{(color online). Schematic of the optical pump / multi-THz probe
setup. The output of the amplified Ti:sapphire laser system is split into three branches: (1) The optical pump pulse with variable delay $\tau$;
(2) the THz probe pulse with variable delay $T$; and (3) the gating pulse for field-resolved electro-optic detection. The sample is mounted in a
liquid helium cryostat equipped with CVD diamond windows.} \label{fig:setup}
\end{figure}

Our multi-THz setup is based on a home-built low-noise Ti:sapphire amplifier system for intense 12-fs light pulses centered at a photon energy
of $1.55~\mathrm{eV}$ (800~nm wavelength).\cite{Huber03} Pulse energies up to 0.5 $\mu$J are generated at variable repetition rates of up to
4~MHz. The simplified schematic of the optical pump$~$/$~$multi-THz probe setup is depicted in Fig.~\ref{fig:setup}. Part of the laser output
denoted as branch 1 is used for optical pumping of the sample. The beam in branch 2 generates phase-locked multi-THz probe transients by optical
rectification in a $50~\mathrm{\mu m}$ thick GaSe crystal.\cite{Huber00} A pair of parabolic mirrors focus the THz beam to the spot excited by
the pump beam. The multi-THz transients are focused onto the sample which is held at a preset substrate temperature $T_\mathrm{L}$. The
transmitted radiation is refocused by another set of two parabolic mirrors onto the $50~\mathrm{\mu m}$ GaSe electro-optic
sensor.\cite{Kubler04} By varying the delay \texttt{T} of the gating pulse (branch 3) with respect to the transient, the time evolution of
electric field can be traced for selected delay times $\tau$ of the pump pulse. Fourier analysis of the time-domain data yields two-dimensional
(2D) maps of both absolute amplitude and phase of the pump-induced transmission change. The time resolution along the $\tau$-axis in our setup
is determined by the detector bandwidth and set to be approximately $40~\mathrm{fs}$. The real and imaginary parts of the refractive index of
the VO$_2$ thin film are extracted from the measured equilibrium transmission and the pump-induced transmission change by numerical solution of
a corresponding Fresnel equation.\cite{Huber01} Any other optical constants, e.g. optical conductivity $\sigma(\omega,\tau)$ or permittivity
$\epsilon(\omega,\tau)$, are available from these data through basic electrodynamic relations.\cite{Jackson98}

The spectrally-integrated dynamics of the optical conductivity is measured if the delay \texttt{T} is fixed at the maximum of the multi-THz
transient and only the delay of the pump pulse $\tau$ is scanned. Since the total spectral intensity of the probe pulse is proportional to the
peak value of the electric field, a measured pump-induced change of its amplitude characterizes changes in absorption or optical conductivity of
a sample in the whole spectral range covered by the multi-THz transient.

\section{Results}

\subsection{Equilibrium optical response of the VO$_2$ sample}

\begin{figure}
\centerline{\includegraphics[angle=0,width=1\columnwidth]{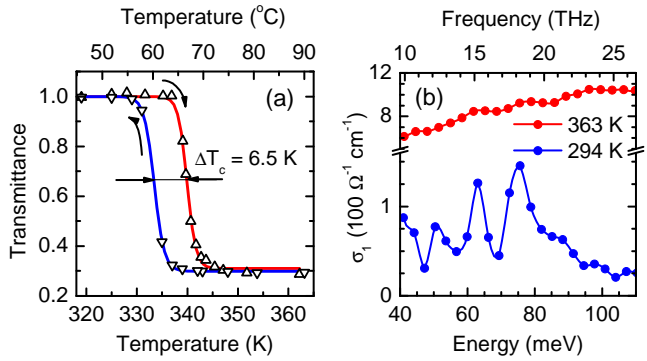}} \caption{(color online). (a) Temperature dependence of the normalized
transmittance through a VO$_2$ thin film upon heating (red guide to the eye) and cooling (blue guide to the eye). (b) Real part of the optical
conductivity of the VO$_2$ sample below and above $T_\mathrm{c}$.} \label{fig:equilibrium}
\end{figure}

When the sample temperature is varied across the phase transition, the transmission in the THz range shows a pronounced hysteresis illustrated
in Fig.~\ref{fig:equilibrium}(a). The insulator-metal transition occurs at $T_\mathrm{L} = 339.5\pm 1$~K upon heating and at $333\pm 1$~K when
cooling down. The resulting narrow hysteresis width of 6.5~K attests to the high quality of the thin film sample.\cite{Suh04} The hysteretic
behavior confirms the first-order character of the phase transition in VO$_2$. Previous calorimetric studies demonstrated the presence of a
latent heat of 241~J/cm$^3$ at the transition point.\cite{Berglund69}  All measurements presented in this paper were acquired on the heating
part of the hysteresis cycle.

Fig.~\ref{fig:equilibrium}(b) shows the real part of the optical conductivity $\sigma_1(\omega)$ for two selected sample temperatures below and
above $T_\mathrm{c}$. Although $\sigma_1$ of insulating VO$_2$ is expected to vanish, at $T_\mathrm{L}$ = 295~K it exhibits pronounced maxima at
$\hbar\omega$ = 50, 62, and 74~meV (13, 15, and 18~THz). These peaks correspond to the transverse optical phonon resonances of the monoclinic
lattice related to vibrations of the oxygen cages surrounding the V atoms.\cite{Barker66,Gervais85} Vanadium-dominated normal modes, in
contrast, are known to oscillate in the low-frequency regime between 2 and 6~THz.\cite{Gervais85,Cavalleri04} The spectral region above 85~meV
is free of infrared-active resonances and displays a low conductivity in the insulating phase. The spectral positions and weights of the phonon
resonances in the monoclinic phase are in good agreement with previous infrared studies on VO$_2$ single crystals.\cite{Barker66,Gervais85} The
observed phonon resonances can be identified by comparing their eigenfrequencies and oscillator strengths with data reported by Barker et
al.\cite{Barker66} Two strong resonances at 62 and 74~meV are assigned to the oxygen vibrations perpendicular to the monoclinic $a_M$ axis
(collinear with $c_R$ of the rutile phase). The weaker resonance around 50~meV is related to the vibration polarized along the $a_M$ axis. As
expected for polycrystalline thin films, the phonon linewidths of about 10~meV observed in our sample are about twice larger than those of
single crystalline samples (4-5~meV).\cite{Barker66} First, our thin-film sample are polycrystalline and have reduced long-range order which
causes a broadening of the observed lattice modes. Second, the experiments performed with polycrystalline thin films always average over
different crystal orientations.

The optical conductivity in the metallic phase increases monotonically with temperature. However, it does not increase with decreasing energy as
expected for a Drude response of conventional metals [Fig.~\ref{fig:equilibrium}(b)]. The peculiar spectral shape arises from the co-existence
of insulating and metallic domains typical of a first-order transition. Hence, in a DC resistivity measurement, a sample appears metallic as
soon as a percolation path connecting metallic domains is formed. On the other hand, the optical response averaged over many domains is
determined by an effective dielectric function that depends on the volume fractions of the monoclinic and rutile phase. A good description of
experimental infrared spectra using Bruggeman's effective medium theory\cite{Stroud75} has been demonstrated.\cite{Choi96,Chang05,Jepsen06}
Based on this success, recent studies rely on the effective medium theory for a quantitative analysis of the optical
spectra.\cite{Hilton07,Rozen06,Qazilbash07,Qazilbash09}

\subsection{Time-resolved spectrally integrated multi-THz response of photoexcited VO$_2$}\label{integrated}

\begin{figure}
\centerline{\includegraphics[angle=0,width=1\columnwidth]{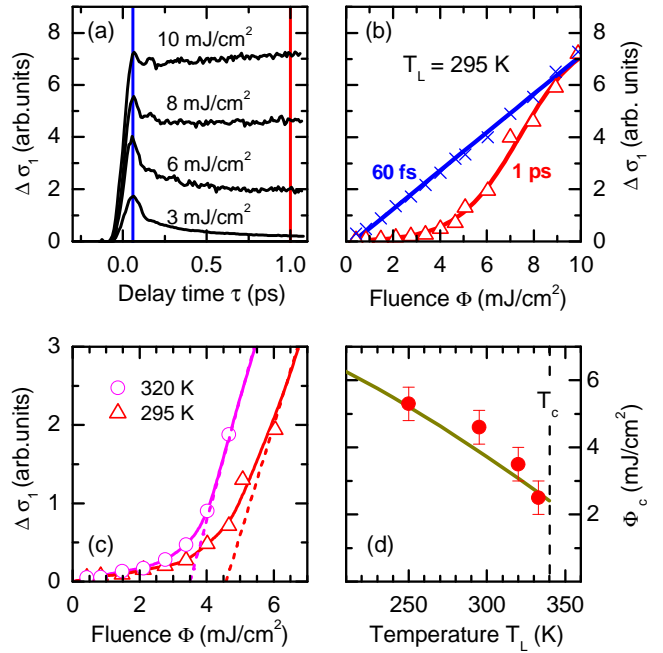}} \caption{(color online). (a) Spectrally integrated transient change of the
THz conductivity after excitation by a 12-fs near-infrared laser pulse at $T_\mathrm{L}$ = 295~K for various pump fluences. (b) Fluence
dependence of $\Delta\sigma_1(\tau)$ at $\tau$ = 60~fs (blue crosses) and 1~ps (red triangles). (c) Extrapolation of $\Delta\sigma_1$(1~ps)
curves (red triangles: 295~K, magenta circles: 320~K) to a critical fluence of $\Phi_c$(295~K) = 4.6~mJ/cm$^2$ and $\Phi_c$(320~K) =
3.5~mJ/cm$^2$, respectively. (d) Dependence of threshold fluence $\Phi_c$ on lattice temperature $T_\mathrm{L}$. The solid line is the
thermodynamic energy difference between the metallic state and the insulating state at given temperature calculated according to
Ref.~\onlinecite{Berglund69}. The dashed line marks the phase-transition temperature $T_\mathrm{c}$.} \label{fig:thresholds}
\end{figure}

Now we consider the non-equilibrium response of VO$_2$ induced by the optical pump pulse. The pump photons at 1.55~eV promote electrons from the
split-off $d_{||}$ band below the Fermi level to the conduction band $\pi$ states [see Fig.~\ref{fig:structure}(b)]. The optically generated
free electron-hole pairs, in turn, give rise to a finite photoconductivity. Fig.~\ref{fig:thresholds}(a) depicts the ultrafast dynamics of the
spectrally integrated conductivity change $\Delta\sigma_1(\tau)$, for a series of pump fluences $\Phi$ recorded at a sample temperature of
$T_\mathrm{L}$ = 295~K. The initial increase of the infrared conductivity peaks within a resolution-limited time interval of 60~fs, marking the
completion of the excitation process. At low excitation densities, a sharp onset of the pump-induced signal is followed by a non-exponential
sub-ps decay. At higher fluences an increasing background of long-lived conductivity appears that is constant within a time window of 10~ps. The
density of directly excited photocarriers contributing to the initial conductivity depends linearly on the fluence. Thus, we expect the
quasi-instantaneous signal to scale linearly with $\Phi$, as well. Extracting the values of the pump-induced conductivity at $\tau$ = 60~fs
confirms this expectation, as shown in Fig.~\ref{fig:thresholds}(b). In contrast, the corresponding value at $\tau$ = 1~ps vanishes for small
fluences while it increases rapidly above a threshold $\Phi_c$(295~K) = 4.6~mJ/cm$^2$, which is determined by extrapolating the linear part of
the latter graph to zero [inset of Fig.~\ref{fig:thresholds}(b)]. The threshold in excitation density separates two regimes: below $\Phi_c$ the
directly photo-generated electron-hole pairs populate delocalized states and thus give rise to the initial photoconductivity. However, the
lifetime of these states is limited by a very effective trapping or relaxation mechanism that results in the rapid sub-ps decay of the
quasi-instantaneous signal. Increasing the excitation density above the threshold of only one absorbed photon per approximately 10 V-V dimers
triggers a cooperative transition from the insulating to the metallic state that renders the relaxation pathway inoperative. The subsequent
persistence of the photoconductivity indicates the transition to the metallic phase. However, the initial insulating state is restored before
the arrival of the next pump pulse due to heat dissipation to the diamond substrate (the substrate temperature $T_\mathrm{L}$ was kept below the
hysteresis region).

\begin{figure}
\centerline{\includegraphics[angle=0,width=1\columnwidth]{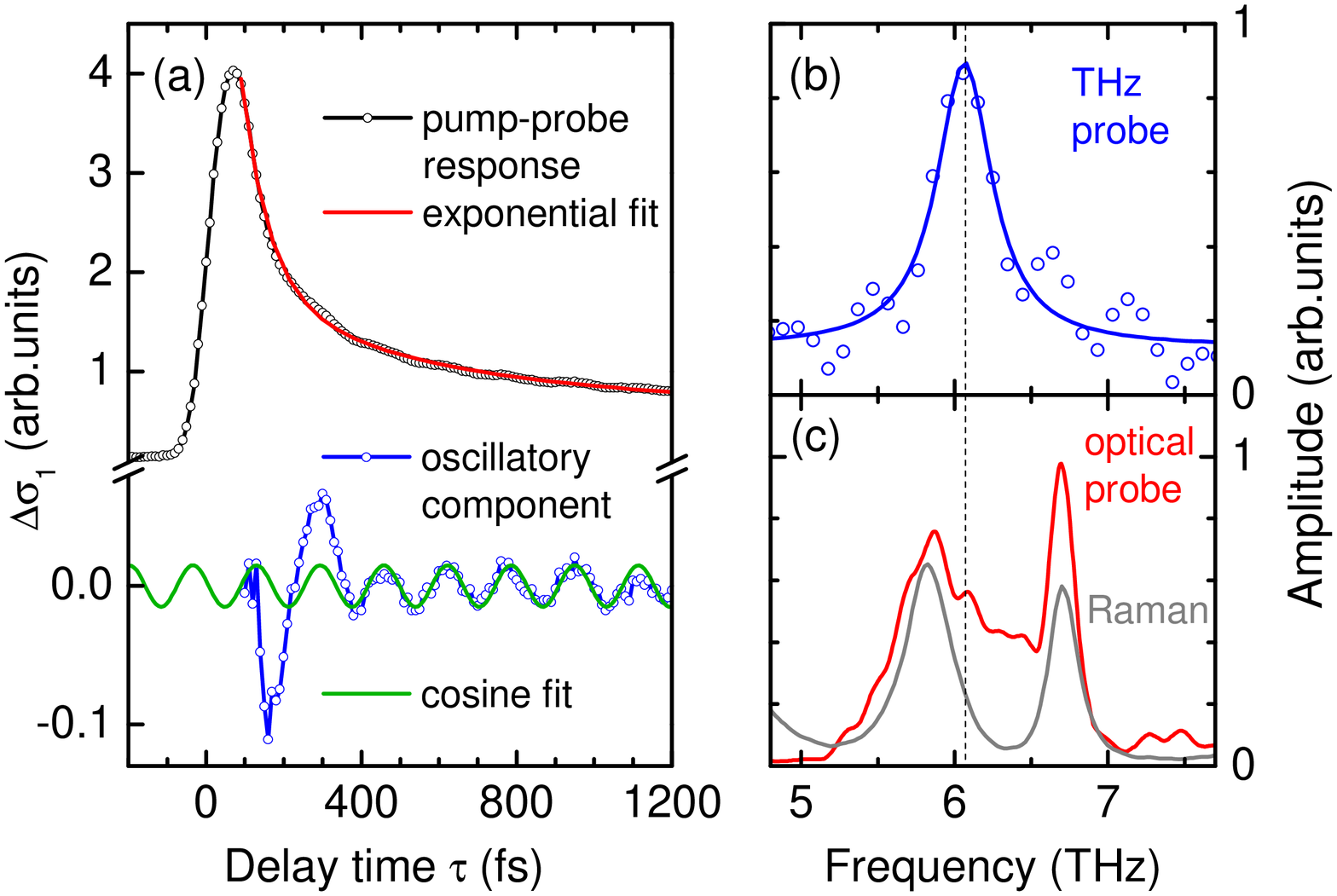}} \caption{(color online). Decomposition of the decay dynamics. (a)
Spectrally integrated dynamics of the THz conductivity after excitation at fluence $\Phi$ = 14~mJ/cm$^2$ and temperature $T_\mathrm{L}$ = 4~K
(black circles). Subtracting the biexponential decay component (red line) reveals a coherent oscillation (blue circles) to which a cosine
function with a frequency of $\omega/2\pi$ = 6.1 THz (green line) was fitted. (b) Spectrum of the oscillatory component (blue circles) and fit
by a Lorentzian function (blue line). (c) The spectrum of the oscillatory modulation observed in a degenerate pump-probe
experiment\cite{Cavalleri04} at 1.55~eV (red line) and the unpolarized spontaneous Raman spectrum\cite{Petrov02} (gray line).} \label{fig:osc1d}
\end{figure}

Raising the sample temperature closer to $T_\mathrm{c}$ leads to a decrease of the threshold fluence $\Phi_c$ [see Fig.~\ref{fig:thresholds}(c)]
and the photoinduced conductivity becomes more long-lived as compared to room temperature. On the other hand, the cooperative transition may be
suppressed when the lattice is cooled to cryogenic temperatures although the excitation density exceeds its critical room-temperature value [see
Sec.~\ref{resolved}]. Indeed, as it is shown in Fig.~\ref{fig:thresholds}(d) the fluence threshold experiences a significant reduction as the
critical temperature is approached from below. Since the threshold fluence corresponds to the minimal energy of the pump pulse necessary to
induce the cooperative transition into the metallic phase, it is instructive to compare it to the difference of the thermal energy of the VO$_2$
lattice between the metallic and insulating phases. This difference can be calculated as the integral of the lattice heat capacity between a
given temperature and $T_\mathrm{c}$ plus the latent heat of the insulator-metal transition. The resulting curve calculated using the data from
Ref.~\onlinecite{Berglund69} is shown in Fig.~\ref{fig:thresholds}(d) as a solid line. In order to enable a direct comparison with pump fluence,
the thermal energy density is recalculated in terms of a surface energy density assuming a 100-nm-thick excited layer. The agreement between
both data sets is remarkable.

Since the electronic contribution to the specific heat in the studied temperature range is negligible compared to the lattice contribution, we
can conclude that optical pumping above $\Phi_c(T)$ should induce a structural transformation of VO$_2$ into the rutile phase on the time scale
shorter than 1~ps after the excitation which we have chosen for the estimate of the excitation threshold. Indeed, the dynamics of the optical
conductivity in Fig.~\ref{fig:thresholds}(a) confirms this assumption.

It should be noted that the agreement between optical pump and thermal energies [Fig.~\ref{fig:thresholds}(c)] does not imply a thermal
character of the photoinduced transition in VO$_2$. Clearly, the sub-ps switching time would be untypically short for a thermally driven
process. As a matter of fact, both coherent (optical switching) and incoherent (thermal heating) mechanisms are observed, but they are well
separated in time. For example, after photoexcitation at 295~K the pump-probe signal starts rising again after several tens of ps and reaches
its maximum level after 300 ps, independent of the excitation fluence [not shown]. The phase growth in the thin film is driven by hot phonons
propagating at the speed of sound and proceeds incoherently at spatially separated sites. The conductivity dynamics of this thermal phase
transition activated by laser heating was studied in detail by Hilton et al.\cite{Hilton07} and is not a subject of this paper.

\begin{figure}
\centerline{\includegraphics[angle=0,width=0.9\columnwidth]{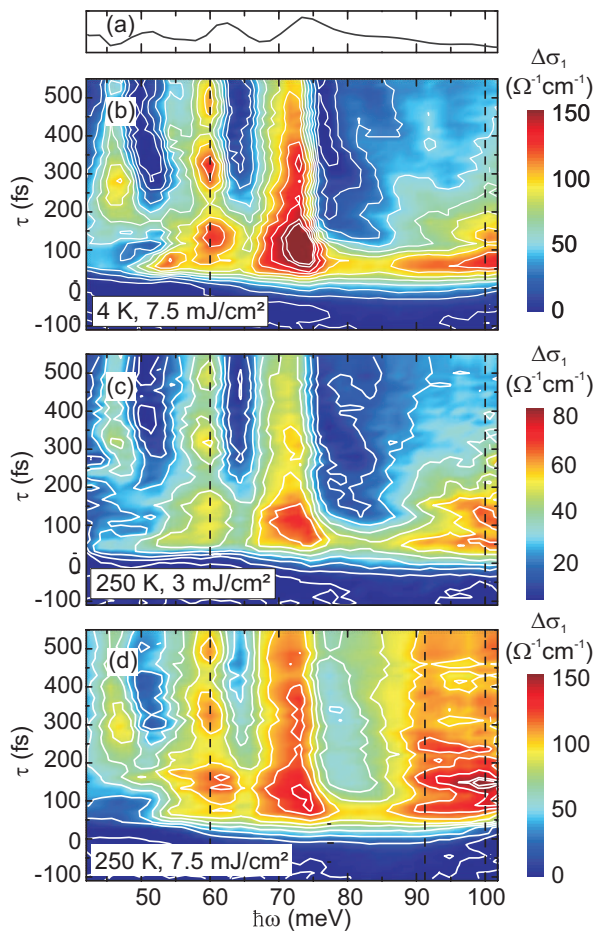}} \caption{(color online). 2D optical pump$~$/$~$multi-THz probe data: (a)
Equilibrium conductivity of insulating VO$_2$ at 295~K. Color plots of the pump-induced changes of the conductivity
$\Delta\sigma_1(\omega,\tau)$: (b) at $T_\mathrm{L}$ = 4~K and an incident fluence of $\Phi$ = 7.5~mJ/cm$^2$; at $T_\mathrm{L}$ = 250~K and pump
fluence (c) $\Phi$ = 3~mJ/cm$^2$ and (d) $\Phi$ = 7.5~mJ/cm$^2$. The broken vertical lines indicate the frequency positions of cross sections
reproduced in Fig.~\ref{fig:1dcuts}.} \label{fig:2dmaps}
\end{figure}

A closer look at the decay dynamics of the THz conductivity reveals a fast oscillation superimposed on the monotonically decreasing signal
[Fig.~\ref{fig:osc1d}(a)]. The presence of this feature in the pump-probe signal is a first substantial signature of the participation of
lattice modes in the phase transition. Fig.~\ref{fig:osc1d}(a) shows how subtraction of a biexponential decay from the pump-probe signal reveals
a coherent cosine-like oscillation with a frequency of 6.1~THz. Apparently, the first cycle of the observed coherent oscillation is notably
stronger and cannot be well described by a simple cosine function. This points out a strongly non-equilibrium character of the multi-THz
response during the first oscillation period of 160~fs when the lattice distortion is accompanied by a strong change in electronic structure.
The following harmonic oscillations describe a pure lattice vibration around a new quasi-equilibrium configuration. It should be noted that
coherent oscillations at a constant frequency of 6.1~THz are observed at all lattice temperatures below $T_\mathrm{c}$ independent of the
excitation fluence. Pumping the metallic phase ($T > T_\mathrm{c}$) reduces the conductivity and the sign of the spectrally integrated response
is inverted. Thus, for delay times $\tau > 100$~fs, the observed behavior is typical of a hot electron gas.

The Fourier transform of the oscillatory component excluding the first cycle is shown in Fig.~\ref{fig:osc1d}(b). It is well fitted by a single
Lorentzian term. In contrast, all-optical ultrafast reflectivity experiments\cite{Cavalleri04} have revealed impulsive excitation of two
vibrational modes at 5.85~THz and 6.75~THz which coincide with totally symmetric $A_g$ Raman modes of the monoclinic
lattice\cite{Petrov02,Schilbe02} [Fig.~\ref{fig:osc1d}(c)]. These phonon modes are critical to the metal-insulator transition in VO$_2$ and
describe stretching and tilting of vanadium dimers which map the monoclinic onto the rutile lattice.\cite{Paquet80} Thus, the corresponding
resonances are only observed in the ground insulating state of VO$_2$ probed by spontaneous Raman spectroscopy. The observation of a single
oscillation frequency instead of the two frequencies in the dimerized monoclinic phase indicates the higher lattice symmetry of the excited
state probed in our experiment.

In general, pump-probe techniques utilizing an impulsive excitation may be sensitive to ground-state as well as to excited-state
vibrations.\cite{Yan85,Dexheimer00} This can be clearly seen in Fig.~\ref{fig:osc1d}(c) where the experimental data from a degenerate
femtosecond pump-probe experiment at a photon energy of 1.55~eV are shown.\cite{Cavalleri04} Besides two $A_g$ Raman modes of the ground state
of VO$_2$ an additional shoulder appears between both maxima, exactly at the frequency of 6.1~THz where the coherent oscillation in the
multi-THz probe occurs (see vertical dashed line in Fig.~\ref{fig:osc1d}). In contrast, the multi-THz probe obviously is not sensitive to the
coherent wave packet dynamics in the ground state of VO$_2$ and probes almost exclusively the coherent oscillation in the excited electronic
state. The reason for such selectivity cannot be understood based solely on the spectrally integrated data which do not unravel the multiple
participating degrees of freedom. This limitation is overcome in a full 2D optical pump$~$/$~$multi-THz probe experiment.

\begin{figure}
\centerline{\includegraphics[angle=0,width=1\columnwidth]{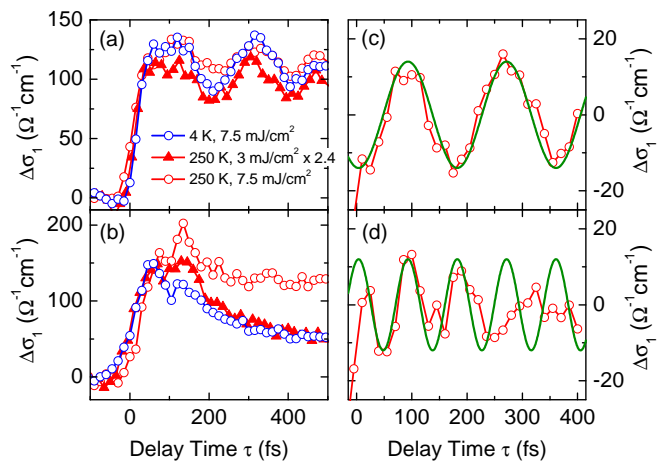}} \caption{(color online). Cross sections of Fig.~\ref{fig:2dmaps} along the
time axis $\tau$ for a photon energy of (a) $\hbar\omega$ = 60~meV and (b) $\hbar\omega$ = 100~meV. The curves taken at $\Phi$ = 3~mJ/cm$^2$ are
scaled up by a factor of 2.4. (c) and (d) The oscillating components of the cross sections through the 2D scan in Fig.~\ref{fig:2dmaps}(d) for a
photon energy of (c) $\hbar\omega$ = 60~meV and (d) $\hbar\omega$ = 92~meV. Green solid lines: (c) fit of the oscillating component by a cosine
function with frequency of 6.1~THz, (d) the fitting function shown in panel (c) squared with subtracted constant background. The oscillation
occurs at doubled frequency of 12.2~THz.} \label{fig:1dcuts}
\end{figure}

\subsection{Spectrally resolved dynamics}\label{resolved}

The spectrally resolved pump-probe response is obtained by measuring the complete time profiles of multi-THz transients as well as their
pump-induced changes at varying pump-probe delay times $\tau$ and subsequent Fourier analysis as described in Sec.~\ref{experiment}. The
resulting 2D plots shown in Fig.~\ref{fig:2dmaps} depict the spectral profiles of the conductivity changes $\Delta\sigma_1(\omega,\tau)$ as a
function of time delay $\tau$ at different temperatures and pump fluences. The equilibrium optical conductivity spectrum of the insulating phase
of VO$_2$ is depicted in Fig.~\ref{fig:2dmaps}(a) for comparison. Figs.~\ref{fig:2dmaps}(b) and (c) show data for an excitation fluence $\Phi$
below the transition threshold $\Phi_c$. As discussed in Sec.~\ref{integrated} the threshold fluence depends on temperature $T_\mathrm{L}$. In
particular, excitation densities that completely switch the VO$_2$ to the metallic phase at room temperature are not sufficient to do so if the
lattice is cooled down to 4~K. Performing the 2D measurement with a fluence of $\Phi$ = 7.5~mJ/cm$^2$, at $T_\mathrm{L}$ = 4~K, results in the
data displayed in Fig.~\ref{fig:2dmaps}(b). The observed dynamics are essentially identical to below-threshold excitation with the fluence
$\Phi$ = 3~mJ/cm$^2$ at $T_\mathrm{L} = 250$~K shown in Fig.~\ref{fig:2dmaps}(c).

Within our time resolution of 40~fs, ultrafast photo-doping induces a quasi-instantaneous onset of conductivity due to directly injected mobile
carriers. For pump fluences below the threshold, the electronic part of the optical conductivity $\Delta\sigma_1(\omega,\tau)$ decays promptly
within approximately 400~fs. In contrast, the pump-induced changes in the phonon resonances are more long-lived. While the onset of the phononic
response varies for the three modes, we find intriguing common features. Photoexcitation induces increased polarizability on the low-frequency
side of each phonon resonance while minimal change is seen on the blue wing: the resonance frequencies are red-shifted. For all modes, the
change in frequency is superimposed on a remarkable coherent oscillation of $\Delta\sigma_1(\omega,\tau)$, along the pump-probe delay axis
$\tau$. This phenomenon is most notable at a THz photon energy of 60 meV marked by a vertical broken line in Fig.~\ref{fig:2dmaps}. The
corresponding cross sections of the 2D maps are shown in Fig~\ref{fig:1dcuts}(a).

After subtraction of a slowly varying background, the coherent oscillation is well fitted by a cosine function with a frequency of 6.1~THz, as
shown in Fig~\ref{fig:1dcuts}(c). Thus, the true origin of the periodic conductivity modulation that appeared already in the spectrally
integrated data of Fig.~\ref{fig:osc1d} is related to the periodic change in the phononic response. The analysis shows that the failure of the
simple decomposition in explaining the spectrally integrated conductivity dynamics for $\tau < 400$~fs [first oscillation cycle in
Fig.~\ref{fig:osc1d}(a)] arises from the presence of the quickly decaying electronic contribution in the total response.

As argued in Sec.~\ref{integrated}, the coherent oscillations at the single frequency of 6.1~THz observed by multi-THz probing in contrast to
the all-optical experiment\cite{Cavalleri04} [Fig.~\ref{fig:osc1d}] indicate that the utilized multi-THz probe is more sensitive to the coherent
dynamics of the excited electronic state of VO$_2$. The 2D maps clearly demonstrate that the origin of the coherent dynamics traced by the
multi-THz probe originate in the \emph{anharmonic} coupling between the high-frequency oxygen and low-frequency vanadium lattice vibrations.
Since the multi-THz response is dominated by the vanadium vibration inherent to the rutile phase, we assume that the anharmonicity of the VO$_2$
lattice in the excited electronic state is strongly enhanced as compared to the insulating ground state.

The spectral response of VO$_2$ and its dynamics shown in Figs.~\ref{fig:2dmaps}(b) and (c) are very similar. This fact once more convincingly
demonstrates the interchangeability of temperature and fluence. We may thus conclude, that the observed interplay of electronic and phononic
degrees of freedom that eventually leads to the emergence of the metallic phase, is not unique to the optically driven case but plays an
important role in the thermal transition as well.

Fig.~\ref{fig:2dmaps}(d) shows the 2D map measured with an excitation fluence of $\Phi$ = 7.5~mJ/cm$^2$ exceeding a threshold fluence $\Phi_c$ =
5.3~mJ/cm$^2$ at a temperature $T_\mathrm{L} = 250$~K. The spectral response of the optical phonon resonances is qualitatively similar to
Figs.~\ref{fig:2dmaps}(b) and (c) with the same dynamics characterized by the coherent oscillations [see Fig.~\ref{fig:1dcuts}(a)]. In contrast,
the dynamics of the electronic photoconductivity differs profoundly: After a resolution-limited onset of $\Delta\sigma_1(\omega,\tau)$ due to
nearly instantaneous photodoping, the THz conductivity levels off at a high value, indicating the final transition into a metallic phase. This
behavior is seen also in Fig.~\ref{fig:1dcuts}(b) where the cross sections of the 2D maps at the energy of 100~meV are shown.

Remarkably, at this energy there are no pronounced oscillations imprinted on the electronic conductivity, in contrast to the phononic response.
Nevertheless, careful inspection reveals a periodic modulation of the multi-THz response around 90~meV. This energy range is well above the
highest phonon resonance and related to the electronic part of the conductivity. Most clearly, this effect is seen in the cross section of
Fig.~\ref{fig:2dmaps}(d) made at an energy of 92~meV. The oscillating component in this cross section shown in Fig.~\ref{fig:1dcuts}(d) occurs
exactly at twice the V-V stretching frequency of 6.1~THz. Moreover, the first three cycles of the oscillation shown in Fig.~\ref{fig:1dcuts}(d)
are well described by squaring the fitting function of the 6.1~THz oscillation in Fig.~\ref{fig:1dcuts}(c) (and subtracting an offset). For
later delay times $\tau$ the modulation amplitude rapidly decreases and a relation to the coherent wave packet motion cannot be traced any more.
The doubled frequency of the oscillation indicates that the dominant term in the electron-phonon coupling of the 6.1~THz phonon mode is
quadratic with respect to the phonon normal coordinate.

The red-shift of the oxygen modes is equal in all three investigated cases and hence depends neither on fluence nor temperature. The same holds
for the period of the modulation, which always consistently matches the vanadium 6.1~THz mode, as displayed in Fig.~\ref{fig:1dcuts}(a).
Interestingly, the amplitude of the modulation scales with fluence but hardly depends on temperature at all.

\section{Discussion}

\begin{figure}[t]
\centerline{\includegraphics[angle=270,width=1\columnwidth]{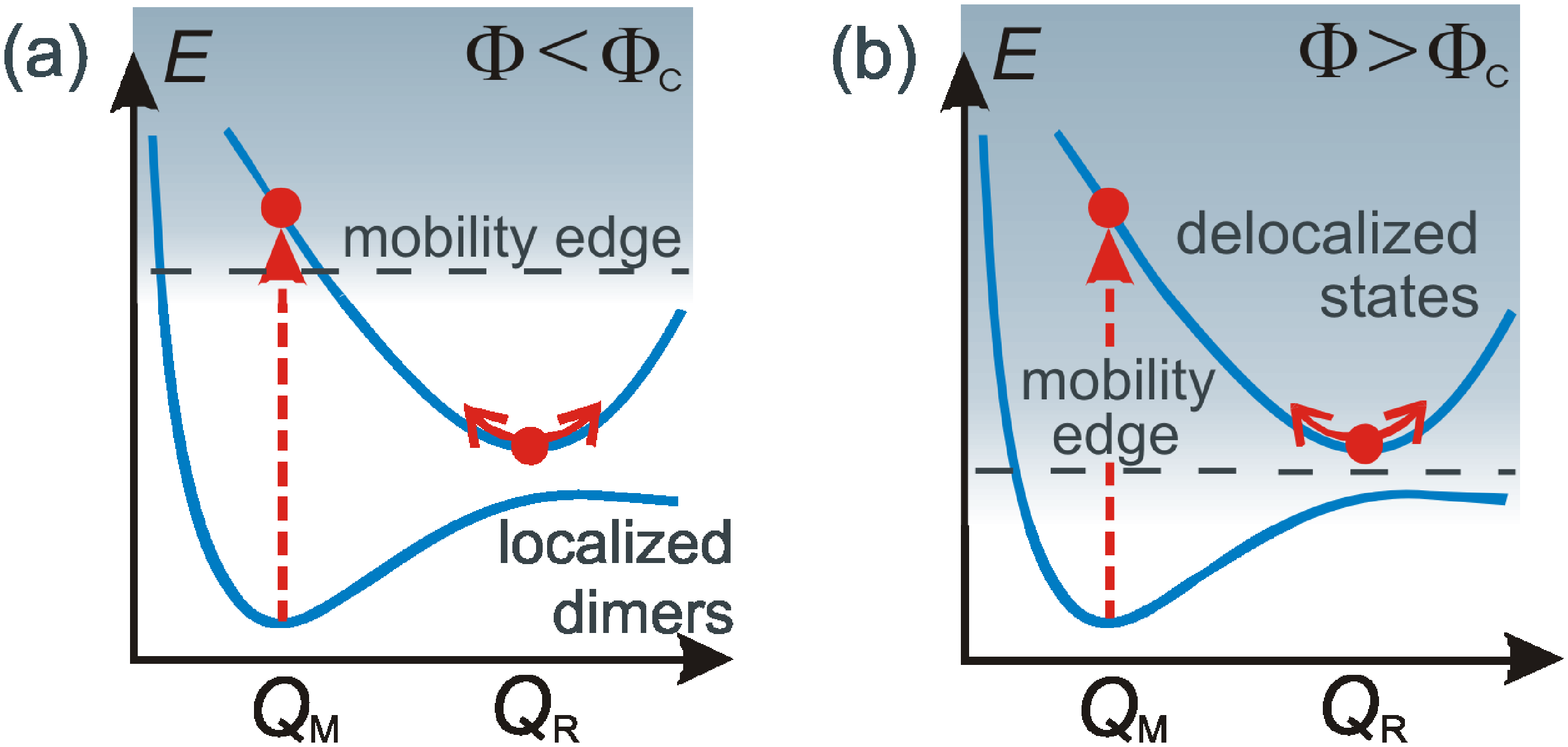}} \caption{(color online). Schematic energy surfaces as a function of a
dimer coordinate $Q$. The energy minima of the potential surfaces $Q_M$ and $Q_R$ correspond to spatial configurations of the V-V dimers in the
monoclinic and rutile phases, respectively. Vertical dashed arrows denote Franck-Condon-type photoexcitation of spin singlets into a conductive
state. (a) Structural relaxation and coherent vibrations about the new energy minimum below the pumping threshold  $\Phi < \Phi_c$. (b) Intense
excitation above the threshold $\Phi > \Phi_c$ leads to a structurally assisted collapse of the mobility edge.} \label{fig:model}
\end{figure}

A qualitative understanding of the experimental observations may be reached using a recent model that describes the electronic structure in
thermal equilibrium within a cluster dynamical mean-field theory.\cite{Biermann05} In this picture, the low-temperature insulating phase is
regarded as a molecular solid of vanadium dimers embedded in a matrix of oxygen octahedra. To first order, the correlated electronic state of
each dimer may be described by bonding and anti-bonding Heitler-London orbitals. The energy dependence of these states on the nuclear positions
(such as V-V separation) is depicted schematically in Fig.~\ref{fig:model}. The minimum of the lower-energy bonding state defines the atomic
position in the monoclinic phase.

Absorption of a near-infrared photon removes an electron from the bonding orbital, destabilizing the dimer, while the lattice site is left in a
vibrationally excited Franck-Condon state. Symmetry requires that the energy minimum of anti-bonding orbitals should be located near the rutile
configuration. Ultrafast photoexcitation thus launches a coherent structural deformation of excited dimers followed by oscillations of $A_g$
symmetry around the new potential minimum (Fig.~\ref{fig:model}). The appreciable lattice deformations impose strong distortions on the
surrounding oxygen octahedra and shift their resonance frequencies. A similar effect has been reported in carbon nanotubes.\cite{Gambetta06}
Remarkably, the frequency shift is very pronounced for the oxygen vibrations polarized perpendicular to the $a_M$ axis (modes at 62 and 74~meV)
and relatively weak for the phonon mode at 50~meV polarized along the $a_M$ axis. This fact attests to an anisotropic character of the
anharmonic coupling between normal modes of the vanadium and oxygen sublattices.

Since the oscillation period of the oxygen-related phonon modes is shorter by a factor of three as compared to the inverse frequency of the
$A_g$ vibration, the eigenfrequencies may follow the structural changes of the vanadium dimer adiabatically. Our experiment directly
demonstrates the influence of a coherent lattice motion on other phonon resonances: An overall red-shift of the oxygen modes attests to a
modified average structure of the vanadium dimers, while the coherent modulation at 6 THz reflects large-amplitude oscillations about the new
potential minimum in the excited state. Although additional shifts of infrared-active phonon modes may be expected from screening via
delocalized charge carriers, the distinctly different decay times of electronic and phononic conductivity prove that this effect is not dominant
[Fig.~\ref{fig:1dcuts}]. The density of excited dimers scales linearly with the laser fluence and so does the amplitude of the observed coherent
modulation. On the other hand, the lattice temperature has no measurable influence on the coherent modulation of the oxygen modes.

The model also provides an instructive explanation of the transient electronic conductivity. Photodoping leaves the dimers initially in a
higher-energy state with delocalized electronic character (Fig.~\ref{fig:model}) generating a quasi-instantaneous onset of electronic
conductivity. Structural distortion drives the dimers into a new energy minimum. The rapid non-exponential decay of the electronic conductivity
seen in Fig.~\ref{fig:1dcuts}(b) indicates that the structural deformation shifts the energy of excited dimers below a mobility edge for
extended electronic states [Fig.~\ref{fig:model}(a)]. This process is analogous to a self-trapping of excitons considered by Dexheimer et
al.\cite{Dexheimer00} Although self-trapped states do not contribute to the electronic conductivity, their structural distortions manifest
themselves by the anharmonic shift of oxygen-related phonon modes that persist for at least 1~ps [see Fig.~\ref{fig:osc1d}(a)]. On the other
hand, at high enough pump fluences electronic conductivity does not decay rapidly. We suggest long-range cooperative effects to account for this
finding: Strong distortions that locally map onto the rutile lattice are expected to adiabatically renormalize the electronic
bands\cite{Chollet05} and lower the mobility edge. Due to the enhanced critical fluctuations at elevated temperatures the system is more
susceptible to this renormalization in accord with experimentally observed lowering of the critical fluence upon approaching $T_\mathrm{c}$.

Instead of the persistent coherent modulation at the frequency of the $A_g$ vibration, a weak modulation of the electronic conductivity at the
doubled frequency is observed. The amplitude of this oscillation rapidly vanishes after about three oscillation cycles
[Fig.~\ref{fig:osc1d}(d)]. This behavior suggests that the lowest-order term in the coupling between the $A_g$ mode and the electronic
conductivity band of the photoexcited state is proportional to $\Delta Q^2$, where $\Delta Q = Q - Q_R$ denotes the deviation of the dimer
coordinate from the metastable rutile structure (see Fig.~\ref{fig:model}). Since, in a first approximation, this coupling is rather weak, the
electronic system may be considered as decoupled from the ionic motion in the non-equilibrium regime after the photoinduced insulator-metal
transition. Consequently, the long-lived high value of electronic conductivity cannot originate from the continuing motion of the V-V dimers.
Instead, we propose to explain this feature by the collapse of the mobility edge\cite{Anderson72} below the Fermi energy
[Fig.~\ref{fig:model}(b)]. In this picture, the critical pump fluence $\Phi_c$ determines whether self-trapping or cooperative delocalization of
electron-hole pairs prevails.

\section{Conclusion}

The ultrabroadband THz experiments presented here allow us to directly trace the the temporal evolution of the electronic conductivity and the
optical phonon resonances during the transient formation of the metallic phase in VO$_2$. This study reveals fingerprints of a coherent V-V
intradimer wave packet motion at 6.1 THz which couples anharmonically to infrared-active phonon modes. The coherent response of the crystal
lattice is explained within a model description of local V-V dimers photoexcited into an antibonding state. The electronic conductivity shows a
weak quadratic coupling to the coherent wave packet motion. A threshold fluence of the excitation pulse for the photoinduced transition is found
to decrease as the equilibrium lattice temperature $T_\mathrm{L}$ approaches the transition temperature $T_\mathrm{c}$. Below threshold, the
electronic correlations remain undisturbed and the mid-infrared conductivity vanishes on a sub-picosecond time scale. Above a critical
excitation density, a cooperative insulator-metal transition occurs on a timescale set by the first half of the oscillation cycle of the V-V
coherent wavepacket motion in the excited state. In so far, the data indicate a strong structural component of the phase transition mechanism.
However, right after the transition to the metallic phase the electronic system is rendered insensitive to the continuing coherent wave packet
motion. Thus the stabilization of the metallic phase must originate from a different mechanism, most likely a collapse of the mobility edge due
to the cooperative modification of the electronic band structure.

\section{Acknowledgements}
We thank I. Perakis, J. Kroha, Th. Dekorsy, A. Cavalleri and S. Wall for stimulating discussions and support. Research at Vanderbilt University
is supported by the National Science Foundation (DMR0210785). R. F. H. and R. H. gratefully acknowledges support from the Alexander von Humboldt
Foundation and Deutsche Forschungsgemeinschaft via Emmy Noether grant HU-1598, respectively.

\end{document}